\documentclass{appolb}
\usepackage{graphicx}
\usepackage{braket}

\begin{document}
\title{Measuring intensity interference in a low multiplicity system $\pi \pi x$ with a new observing method %
}
\author{Q. He, X. He, T. Li
\address{Department of Nuclear Science and Engineering, Nanjing University of Aeronautics and Astronautics (NUAA), 29 Yudao St., Nanjing 210016, China
\\heqh@nuaa.edu.cn
}
\\
}
\maketitle
\begin{abstract}
Prior proposed observing approaches using event mixing technique for Bose-Einstein correlations (BEC) measurements in exclusive reactions with very low multiplicities are still unsatisfactory due to the problems of sample reduction and introducing extra and unnecessary fitting parameters. We here propose an event mixing method with a new mixing cut, named energy sum range (ESR) cut, to investigate two-pion Bose-Einstein correlations (BEC) in reactions with only two identical pions among three final state particles. This mixing method employs two-pion energy sum characteristic to control the mixing procedures, with no requirement on eliminating any original events. Numerical simulations are performed to show the viability of this new BEC observing method.
\end{abstract}
\PACS{25.20.Lj, 14.20.Dh, 14.20.Gk.}
  
\section{Introduction}	

Intensity interference between identical bosons, generally known as Bose-Einstein correlations \cite{Weiner1999,Boal1990,Goldhaber1959,Alexander2003}, is widely used to provide insights into the dynamics process and space-time structure of the particle emitting source created via hadron collisions \cite{Aihara1985,Althoff1986,Choi1995,Decamp1992,Abreu1992,Behrend1990,Korotkov1993,Arneodo1986,Adamus1988,Bailly1989,Adloff1997} or heavy-ion collisions \cite{Chacon1991,Heinz1999,Abelev2006,Aamodt2011PLB,Podgoretsky1989, Lednicky2009, Pratt1984, Lisa2005}. Using this method to measure the spatial size of nucleon resonances excited by hadronic or electromagnetic probes in the non-perturbative QCD energy region (assuming these resonances decay via emitting identical bosons, e.g. $\gamma p \to N^* \to \pi^0\pi^0p$), is however challenging because proper BEC observing method at low energies with low multiplicities is still lacking. 

One key technique involved in observing intensity correlations is how to construct a valid reference sample for measuring the correlation function which is constructed from the detection probabilities $I_1$ and $I_2$:
\begin{equation}
g^{(2)}(p_1,p_2)=\frac{ \braket {I_{12}} }  {\braket {I_1}\braket{I_2}}
\label{2ndInterference}
\end{equation}
where $\braket {I_{12}}$ is the joint probability for the emission of two identical bosons with momenta of $p_1$ and $p_2$, respectively, subject to Bose-Einstein symmetry (BES), while $\braket {I_1}\braket{I_2}$ corresponds to the emission probability in the absence of BES and generally is known as "reference sample". If a Gaussian density profile of the boson-emission source is assumed, Eq. (1) is written as
\begin{equation}
g^{(2)}(p_1,p_2)=g^{(2)}(Q)=N(1+\lambda_2e^{-r_0^2Q^2}),
\label{diseqn}
\end{equation}
where $N$ is the normalization factor, $Q$ the relative momentum of two bosons defined by $Q^2=-(p_1-p_1 )^2$, and $r_0$ the Gaussian radius of the source. The parameter $\lambda_2$ is introduced as a measure of the BEC strength ranging from 0 to 1, where 0 and 1 correspond to completely coherent and totally chaotic emission, respectively. 

A primary method for reference sample construction is the event mixing technique\cite{Kopylov1972, Kopylov1975}, which produces "un-correlated" sample from original sample through making artificial events by randomly selecting two bosons' momenta from different original events. The event mixing method works well for BEC observations in high-energy reactions with sufficiently large multiplicities\cite{Alexander2003}. However, its applications in exclusive reactions with very low multiplicities is still a big challenge. The main reason is that the event mixing is strongly obscured by non-BEC factors such as global conservation laws and decays of resonances\cite{Klaja2010,Aamodt2011}. Conservation laws induce significant kinematical correlations between final states particles and complicate the BEC analysis\cite{Chajecki2008,He2014}.

To develop a proper event mixing method for observing particle correlations in exclusive reactions with low multiplicies, one needs to explore appropriate constraints to manage the mixing process to ensure that the produced reference sample is free of BEC effects but preserves all other kind of correlations arising from global conservation laws and decays of resonances. In the work of Ref. \cite{He2016} they tried to develop an event mixing method for $\pi\pi$ BEC observations in a three final state system $\pi\pi X$, taking the reaction $\gamma p \rightarrow \pi^{0}  \pi^{0} p$ at incident photon energies around 1 GeV (a non-QCD region) as an example. The effects of kinematical correlations due to energy-momentum conservation was investigated, and an event mixing method which contains two mixing cuts was proposed. The first cut, named missing mass consistency (MMC) cut is adopted to conserve the energy momentum of the mixed events and to make them physically meaningful as the original events. The second cut, named pion energy (PE) cut, is used to adjust the slope of the correlation function to extract correct BEC parameters. Although this mixing method works well for $\pi\pi X$ events in pure phase space, a disadvantage of the PE cut is that it needs to eliminate a large portion (about 40\%) of original events and hence reduces the statistics. In order to solve the sample reduction problem, two new mixing methods, energy sum order and invariant-mass/energy hierarchy correspondence cuts, were proposed later \cite{He2018_1,He2018_2} to replace the PE cut. Although the new proposed methods solved the problem of sample reduction,  they introduced extra and unnecessary fitting parameters, and hence leads to a worse analysis accuracy.

In this work we propose a new mixing cut constrraint, named energy sum range (ESR) cut, in order to solve both the sample reduction and extra fitting parameters issue.  This new cut employs the two-pion energy sum to control the mixing procedure. Numerical tests using  $\gamma p \rightarrow \pi^{0}  \pi^{0} p$ events are performed to test the ability of this cut to observe BEC effects. 

\section{Event mixing with ESR constraint}
We search for suitable cut conditions in event mixing according to such a criterion that the cut should affects correlations arising from energy-momentum conservations so strongly that the mixed events still retain the original pure phase space distribution but on the other hand it should be weakly sensitive to BEC correlations so that no BEC correlations in the mixed sample are preserved. By investigating several cut conditions, the two-boson energy sum, $E_{sum}$, is selected empirically as a cut condition in event mixing. Because in the low energy case, the fitting range of the BEC correlation function is very limited, a flat background correlation function from appropriate mixing cut conditions is required for extraction proper BEC parameters. 

The new event mixing method is composed of two constraints. The MMC cut\cite{He2016} is still included in the mixing method, which requires  $|m_X^{mix}-m_X^{ori} |<M_{cut}$, where $M_{cut}$ is the cut window, and $m_X^{mix}$ and $m_X^{ori}$ are the missing particle mass for the mixed event and that for the original event, respectively, in order to force the mixed events to be physically equal to original events and to be located in the allowed phase space region. In addition to the MMC cut, a new constraint, named energy sum range cut (ESR), is introduced. It requires that two events can be mixed only when the following relation is satisfied:

\begin{equation}
 |E_{sum}^{(ori,1)}-E_{sum}^{(ori,2)}|<E_{cut}, and  
\label{eqnESR}
\end{equation} 
\begin{equation}
    min(E_{sum}^{(ori,1)},E_{sum}^{(ori,2)})<E_{sum}^{mix}<max(E_{sum}^{(ori,1)},E_{sum}^{(ori,2)}),  
\label{diseqn}
\end{equation} 
where $E_{sum}^{(ori,1)}$ and $E_{sum}^{(ori,2)}$ are the two-boson energy sums in the two original event, $E_{sum}^{mix}$ the two-boson energy sum in the mixed event. $E_{cut}$ is a cut window and its optimum value is determined empirically.


\section{Numerical test}
We here adopt the reaction $\gamma p \rightarrow \pi^{0}  \pi^{0} p$, which has only three final state particles including two identical bosons among them, to demonstrate the event mixing method employing the MMC and ESR constraints. Both pure phase space events and BEC-effect events of the $\gamma p \to \pi^{0}  \pi^{0} p$ reaction are used to validate the effectiveness of the ESR cut. The Monte Carlo events generation employs a ROOT utility named 'TGenPhaseSpace' developed by CERN\cite{rootweb} based on the GENBOD function, which employs the Raubold and Lynch method \cite{James1968} and has already been implemented in the CERN library. Details of generating non-BEC sample can be found in Ref. \cite{He2016}.

\begin{figure}[tbh]
\centerline{
\includegraphics[width=0.55\linewidth]{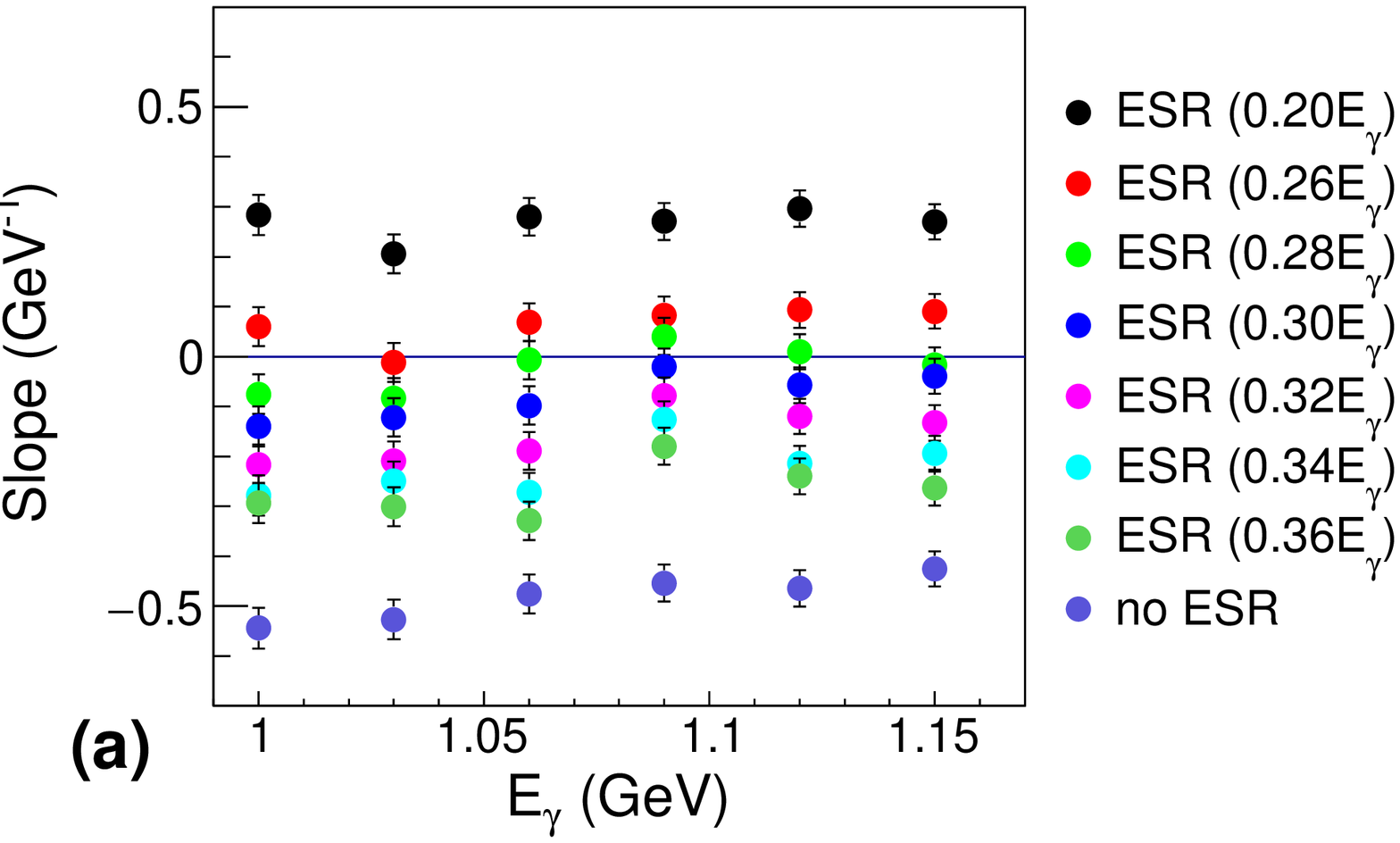}
\includegraphics[width=0.45\linewidth]{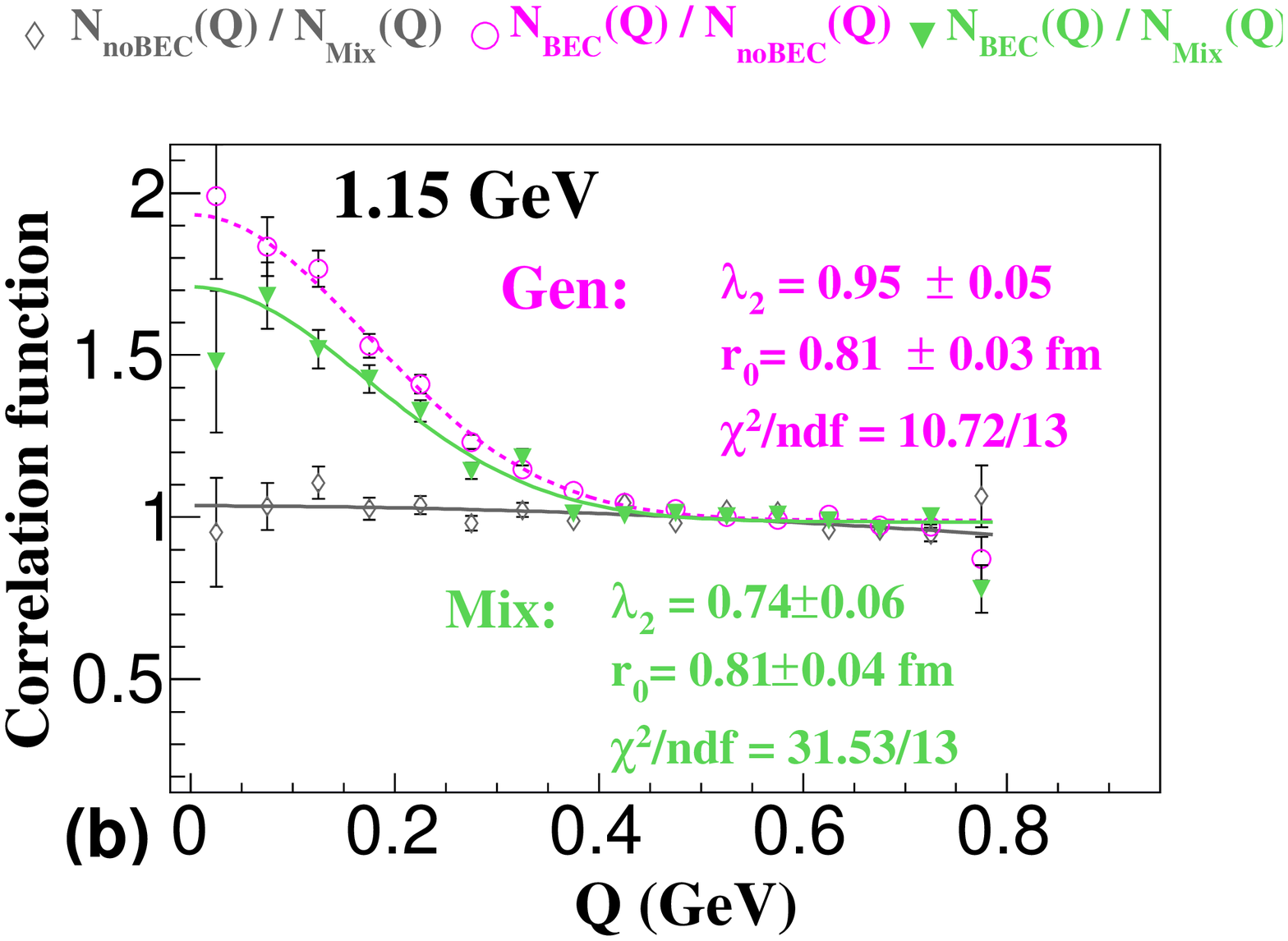}
}
\vspace*{8pt}
\caption{Slope values of the correlation functions obtained with different $E_{cut}$ values for the ESR cut (indicated in parenthesis on the right) at six incident photon energies ($E_{\gamma}$) for the reaction $\gamma p \rightarrow \pi^{0}  \pi^{0} p$. The MMC cut is also used.  b) A typical correlation function obtained by the event mixing with the MMC cut and the ESR cut with $E_{cut}=0.32E_{\gamma}$ for $\gamma p \rightarrow \pi^{0} \pi^{0} p$ events in the presence of BEC effects at incident photon energy of 1.15 GeV. Eq. (2) is used to fit the data to get BEC parameters. \protect\label{slopevseb}}
\end{figure}   

A valid cut should produce a flat correlation function. To satisfy this requirement, the cut window parameter $E_{cut}$ of the ESR cut is adjusted through finding the optimum value which can make a correlation function the closest to a flat line. Six pure phase space $\gamma p \rightarrow \pi^{0}  \pi^{0} p$ event samples free of BEC effects at typical incident photon energies of 1.0, 1.03, 1.06, 1.09, 1.12, 1.15 GeV are generated and used to make mixed sample via the event mixing method using the MMC cut and the ESR cut with different $E_{cut}$ values. Fig. \ref{slopevseb} (a) shows the slope values of correlation functions obtained with different $E_{cut}$ values. The slope value is obtained by fitting a linear function $f(Q)=aQ+b$ to the correlation function. $E_{cut}=0.32E_{\gamma}$ is selected as the optimum value for the ESR constraint owing to a trade-off between flat correlation function for non-BEC sample and effective BEC parameters measurement for BEC sample. With the optimum ESR cut, the correlation function for the non-BEC effect $\gamma p \rightarrow \pi^{0}  \pi^{0} p$ events exhibits  a good flat feature as shown in Fig. \ref{slopevseb} (b).

To investigate the ability of the proposed event mixing method to measure BEC effects, event mixing is also performed for BEC samples of $\gamma p \rightarrow \pi^{0}  \pi^{0} p$ events, which are constructed from the prepared pure phase space samples using the following procedures based on the fact that Eq. (2) has a maximum value $g^{(2)}_{max}=N(1+\lambda_2)$ when $Q$ = 0. The events in the phase space sample is selected to compose the BEC sample when they satisfying the relation  $g^{(2)}(Q)/g^{(2)}_{max}>R$, where $R$ is a random number  uniformly generated in the range from 0 to 1, $Q$ the two pions momentum difference. The probability of $g^{(2)}(Q)/g^{(2)}_{max}>R$ is proportional to $g^{(2)}(Q)$. Thus this method is capable of producing correct density distribution subject to Eq.(2). The BEC parameters for the BEC samples are typically set to be $r_0$=0.8 fm and $\lambda_2$=1.0.
            
With the event mixing method, the BEC effects can be obviously observed in the obtained correlation functions, as shown in Fig.~\ref{slopevseb} (b). For comparison, the ratio of  $Q$ spectrum of the BEC sample to that of the corresponding pure phase-space sample is also shown. It can be seen that the proposed mixing method can reproduce the correlation functions as the input ones. The BEC parameters $r_0$ and $\lambda_2$ are determined by fitting Eq. (2) to the correlation function. Table 1 compares the mixing-obtained BEC parameters with the input ones. It is found that the fit $r_0$ values are in good agreement with the input ones at all energy points, while the  $\lambda_2$ values are a little bit underestimated.

\begin{table}[]
\caption{Comparing input BEC parameters with those from the proposed mixing method.}
\begin{tabular}{c|ccc|ccc}
\hline
\hline
$E_{\gamma}$   & \multicolumn{3}{c|}{Input}           & \multicolumn{3}{c}{Fit}            \\
\cline{2-4} \cline{5-7}
(GeV)& $r_0$ (fm)    & $\lambda_2$   & $\chi^2/ndf$ & $r_0$ (fm)    & $\lambda_2$   & $\chi^2/ndf$ \\
\hline
1.00 & 0.79$\pm$0.02 & 0.97$\pm$0.04 & 9.7/11   & 0.83$\pm$0.04 & 0.79$\pm$0.06 & 30.0/11     \\
1.03 & 0.79$\pm$0.03 & 0.94$\pm$0.05 & 20.0/11   & 0.81$\pm$0.04 & 0.86$\pm$0.06 & 34.8/11     \\
1.06 & 0.80$\pm$0.03 & 0.95$\pm$0.05 & 15.5/12   & 0.86$\pm$0.05 & 0.77$\pm$0.07 & 40.7/12     \\
1.09 & 0.78$\pm$0.02 & 0.98$\pm$0.05 & 10.8/12   & 0.82$\pm$0.04 & 0.78$\pm$0.06 & 29.6/12     \\
1.12 & 0.76$\pm$0.02 & 0.92$\pm$0.04 & 5.3/12   & 0.85$\pm$0.04 & 0.79$\pm$0.06 & 13.3/12    \\
1.15 & 0.81$\pm$0.03 & 0.95$\pm$0.05 & 10.7/13   & 0.81$\pm$0.04 & 0.74$\pm$0.06 & 31.5/13     \\
\hline 
Ave.                  & 0.79$\pm$0.01 & 0.95$\pm$0.02 &         & 0.83$\pm$0.02 & 0.79$\pm$0.02 &   \\
\hline            
\end{tabular}
\end{table}

In order to study the systematic bias introduced by the proposed mixing method, the weighted mean fit values of both $r_0$ and $\lambda_2$ are compared to the weighted mean input values. It is found the mean value $r_0$ (0.83$\pm$0.02) over the six energies is about 5\% overestimated compared to the input one, 0.79$\pm$0.01. The mean value of $\lambda_2$ is found to be 0.79$\pm$0.02, about 17\% underestimated compared to the mean value of the input ones, 0.95$\pm$0.02. 

\begin{figure}[tbh]
\centerline{
\includegraphics[width=0.5\linewidth]{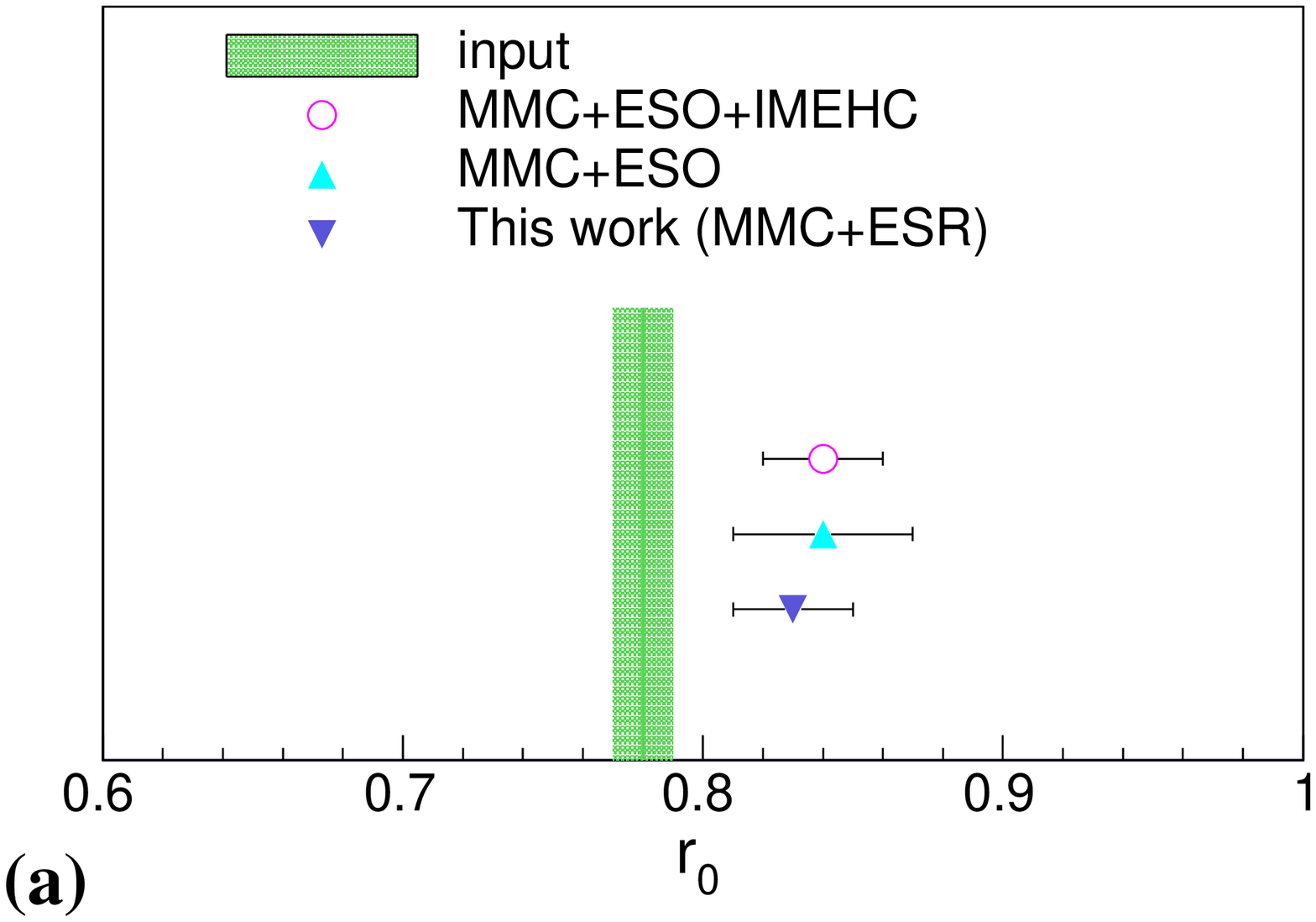}
\includegraphics[width=0.5\linewidth]{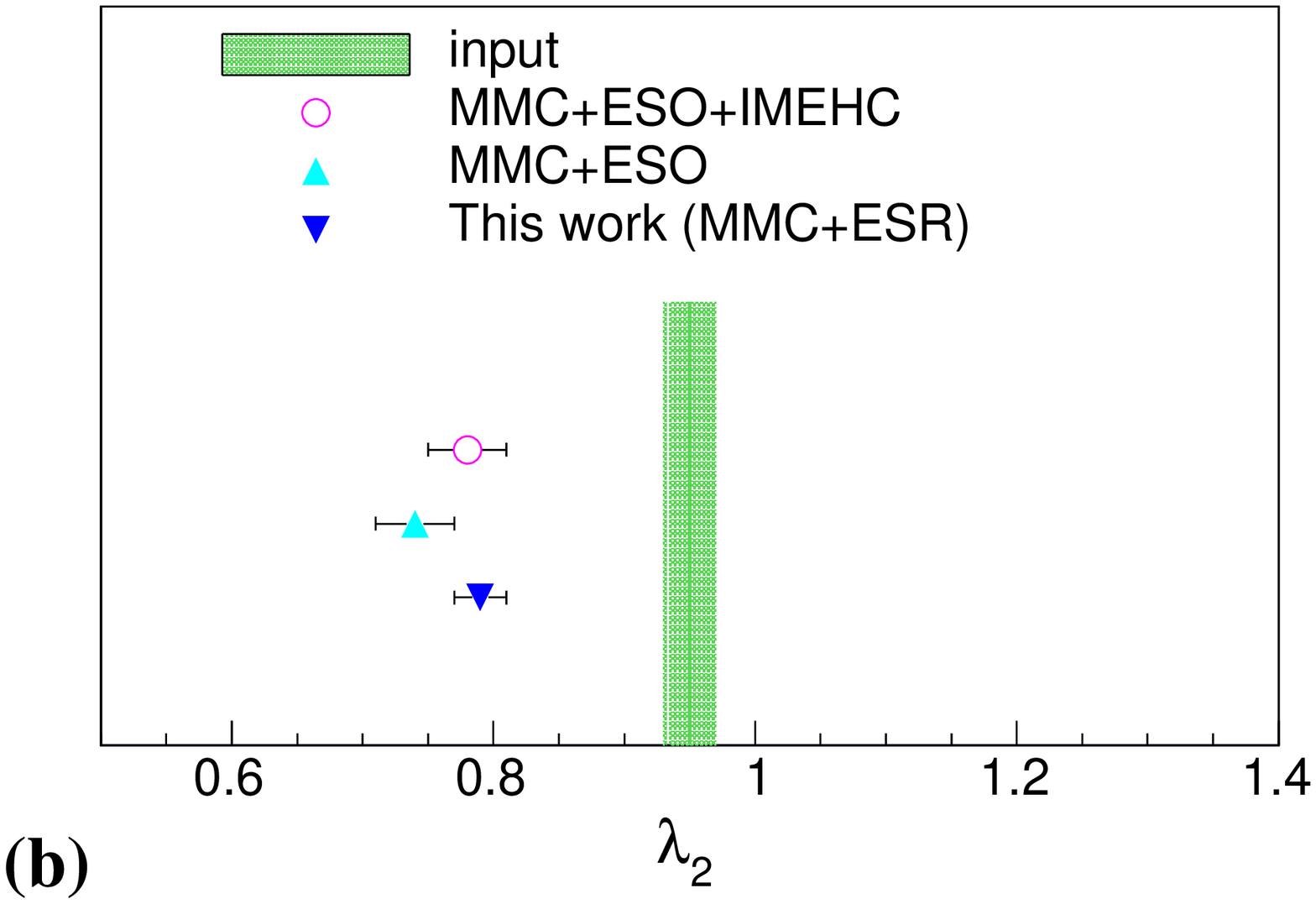}
}
\vspace*{8pt}
\caption{Comparing input BEC parameters with those from the proposed mixing method. For comparison, the results from the other mixing methods are also presented \cite{He2018_1,He2018_2}. \protect\label{rlm}}
\end{figure}

Comparing with two previously proposed mixing methods \cite{He2018_1,He2018_2}, this mixing method induced systematic bias of both $r_0$ and $\lambda_2$ are smaller (Fig. \ref{rlm}), because it avoids the $Q^2$-dependent fitting problem and hence improves the accuracy of the fitting and reduces the systematic bias of the fit BEC parameters. 

Although this new method improves the the accuracy of fitting and reduces the systematic bias of fit BEC parameters, it still introduces systematic bias and in practical applications the BEC parameter obtained by this mixing method should be corrected. Future efforts may focus on improving the systematic bias.

In the experimental side, the two-particle BEC measurement is also affected by the detector acceptance. Generally, such measurements involve a 4$\pi$ detection system. Compared to single particle's detecting efficiency, position resolution, energy resolution and geometric coverage, the joint two-particle detection acceptance may have a stronger impact on the event mixing. A single particle in a mixed event is from real experimental data and its existence is naturally reasonable, but the existence of the two particles in a mixed event may be problematic. This problem is common for GeV-energy particle detection with electro-magnetic calorimeter composed of many independent crystals (e.g. BGO).  An incident energetic particle creates an electromagnetic shower, whose energy deposit commonly extends to several adjacent detector crystals. All corresponding detector crystals are grouped into a cluster. As a result, the detector cannot distinguish two particles if their clusters are overlapping. Although there exist algorithms to separate two overlapping clusters, it may give an ambiguous result. To cope with this situation for the real data BEC analysis, a cluster overlapping cut may be needed, which requires any pair of clusters in an event must share no overlapping crystals both for the real data and mixed events. With this cut, events with two clusters having overlapping crystals are rejected from the event mixing and not counted as original events. Numerical simulations show the correlation function has a steep drop at low $Q$s without the cluster overlapping cut. Generally speaking, for a given detector, specific mixing cut should be taken to eliminate possible detector acceptance impacts on event mixing. A detailed discussion of detection acceptance effects on event mixing is, however, outside the scope of the present paper.

\section{Summary}
A new event mixing method is proposed for two-pion Bose-Einstein correlations (BEC) measurement in reactions with only two identical pions among three final state particles. This mixing method with a new mixing constraint named energy sum range cut eliminates extra and unnecessary fitting parameter and hence improves the systematic bias both for $r_0$ and $\lambda_2$. Numerical simulations with the $\gamma p \rightarrow \pi^{0}  \pi^{0} p$ events at several incident photon energies around 1 GeV are performed to verify this mixing method. It is found that this new method has smaller systematic bias and better fitting uncertainties for both BEC parameters  $r_0$ and $\lambda_2$, compared to two prior proposed mixing methods. In future studies, improvements of the systematic bias are needed.

\section*{Acknowledgments}
This work was supported by the National Natural Science Foundation of China, Grant No. 11805099, and the Fundamental Research Funds for the Central Universities, Grant Nos. NS2018043, 1006-YAH17063


\begin{thebibliography}{0}
\bibitem{Weiner1999} R. M. Weiner, {\it Introduction to Bose-Einstein Correlations and Subatomic Interferometry} (Wiley, Chichester, 1999).
\bibitem{Boal1990} D. Boal, C. Gelbke, B. K. Jennings, {\it Rev. Mod. Phys.} {\bf 62}, 553 (1990).
\bibitem{Goldhaber1959}	G. Goldhaber et al., {\it Phys. Rev. Lett.} {\bf 3}, 181 (1959).
\bibitem{Alexander2003} G. Alexander, {\it Rep. Prog. Phys.} {\bf 66}, 481 (2003).
\bibitem{Aihara1985}	H. Aihara et al., {\it Phys. Rev. D} {\bf 31}, 996 (1985). 
\bibitem{Althoff1986}	M. Althoff et al., {\it Z. Phys. C} {\bf 30}, 355 (1986). 
\bibitem{Choi1995}	S. K. Choi et al., {\it Phys. Lett. D} {\bf 355}, 406 (1995). 
\bibitem{Decamp1992}	D. Decamp et al., {\it Z. Phys. C} {\bf 54}, 75 (1992). 
\bibitem{Abreu1992}	  P. Abreu et al., {\it Phys. Lett. B} {\bf 286}, 201 (1992). 
\bibitem{Behrend1990}	H. J. Behrend et al., {\it Phys. Lett. B} {\bf 245}, 298 (1990). 
\bibitem{Korotkov1993}	V. A. Korotkov et al., {\it Z. Phys. C} {\bf 60}, 37 (1993). 
\bibitem{Arneodo1986}	M. Arneodo et al., {\it Z. Phys. C} {\bf 32}, 1 (1986). 
\bibitem{Adamus1988}	M. Adamus et al., {\it Z. Phys. C} {\bf 37}, 347 (1988).
\bibitem{Bailly1989}	J. L. Bailly et al., {\it Z. Phys. C} {\bf 43}, 431 (1989). 
\bibitem{Adloff1997}	C. Adloff et al., {\it Z. Phys. C} {\bf 75}, 437 (1997).
\bibitem{Chacon1991} A. D. Chacon et al., {\it Phys. Rev. C} {\bf 43}, 2670 (1991). 
\bibitem{Heinz1999}	U. Heinz and B. V. Jacak, {\it Annual Review of Nuclear and Particle Science} {\bf 49}, 529 (1999). 
\bibitem{Abelev2006} B. Abelev et al., {\it Phys. Rev. C} {\bf 74}, 054902 (2006). 
\bibitem{Aamodt2011PLB}	 K. Aamodt et al., {\it Phys. Let. B} {\bf 696}, 328 (2011).
\bibitem{Podgoretsky1989} M. I. Podgoretsky, {\it Fiz. Elem. Chastits At. Yadra} {\bf 20}, 628 (1989); [{\it Sov. J. Part. Nucl.} {\bf 20}, 266 (1989)]. 
\bibitem{Lednicky2009} R. Lednicky, {\it Phys. At. Nucl.} {\bf 67}, 71 (2004).
\bibitem{Pratt1984} S. Pratt, {\it Phys. Rev. Lett.} {\bf 53}, 1219 (1984). 
\bibitem{Lisa2005} M. Lisa, S. Pratt, R. Soltz, and U. Wiedemann, {\it Annu. Rev. Nucl. Part. Sci.} {\bf 55}, 357 (2005).
\bibitem{Kopylov1972}	G. I. Kopylov and M. I. Podgoretsky, {\it Sov. J. Nucl. Phys.} {\bf 15}, 219 (1972).
\bibitem{Kopylov1975} G. I. Kopylov and M. I. Podgoretsky, {\it Sov. Phys. JETP} {\bf 42},  211 (1975).
\bibitem{Klaja2010}	P. Klaja et al., {\it J. Phys. G: Nucl. Part. Phys.} {\bf 37}, 055003 (2010). 
\bibitem{Aamodt2011}	K. Aamodt et al., {\it Phys. Rev. D} {\bf 84}, 112004 (2011). 
\bibitem{Chajecki2008}	Z. Chaj\c ecki and M. Lisa, {\it Phys. Rev. C} {\bf 78}, 064903 (2008). 
\bibitem{He2014}  Q. H. He, Ph.D. Thesis, Tohoku University, 2014.
\bibitem{He2016}  Q. H. He et al., {\it Chin. Phys. C} {\bf 40}, 114002 (2016). 
\bibitem{He2018_1} Q. He, {\it Chin. Phys. C} {\bf 42}, 074004 (2018).
\bibitem{He2018_2} Q. He, {\it Acta Phys. Pol. B} {\bf 49}, 1811 (2018).
\bibitem{rootweb} R. Brun and F. Rademakers, Nucl. Instr. Methods A 389, 81 (1997).
\bibitem{James1968} F. James, Monte Carlo Phase Space, CERN, 68, 15  (1968). 





\end{thebibliography}
\end{document}